\documentclass[aps,pre,reprint,superscriptaddress,letterpaper,twocolumn]{revtex4-1}
\usepackage{graphicx}
\usepackage{dcolumn}
\usepackage{bm}
\usepackage{amsmath}
 \usepackage{amssymb}
\usepackage{CJK}
\usepackage{color}
\newcommand{\nMF}{\textbf{nMF}}
\newcommand{\MLE}{\textbf{MLE}}
\begin{document}

\title{Network reconstruction from asynchronously updated  evolutionary game}

\author{Hong-Li Zeng}
\affiliation{School of Science, and New Energy Technology Engineering Laboratory of Jiangsu Province , Nanjing University of Posts and Telecommunications, Nanjing 210023, China}

\author{Shu-Xuan Wang}
\affiliation{School of Computer Science and Technology, Nanjing University of Posts and Telecommunications, Nanjing 210023, China}

\author{Yan-Dong Guo}
\affiliation{College of Electronic Science and Engineering, Nanjing University of Posts and Telecommunications, Nanjing 210046, China}

\author{Shao-Meng Qin}
 \email{qsminside@gmail.com}
\affiliation{College of Science, Civil Aviation University of China, Tianjin 300300, China}

\date{\today}

\begin{abstract}
The interactions between players of prisoner's dilemma (PD) game are reconstructed with evolutionary game data. All participants play the game with their counterparts and gain corresponding rewards during each round of the game. However, their strategies are updated asynchronously during the evolutionary PD game.
Two inference methods of the interactions between players are derived with naive mean-field (\nMF) approximation and maximum log-likelihood estimation (\MLE) respectively.
The two methods are tested numerically also for fully connected asymmetric Sherrington-Kirkpatrick (SK) models, varying the data length, asymmetric degree, payoff and system noise (coupling strength).
We find that the reconstruction mean square error (MSE) of MLE method is proportional to the inverse of data length and typically half (benefit from the extra information of update times) of that by nMF.
Both methods are robust to the asymmetric degree but works better for large payoff.
Compared with MLE, nMF is more sensitive to the couplings strength which prefers weak couplings.
\end{abstract}

\pacs{Valid PACS appear here}
\maketitle

\section{Introduction}

Networks can be generally used for describing systems composed by interacting elements.
Such systems usually generate big time-series data based on the underlying dynamics or mechanics.
Uncovering the network structures from  observations is crucial to understand the collective dynamics of the systems and has attracted great interest in different fields, ranging from biology system \cite{schneidman2006weak, nguyen2016whole, roudi2011mean, zeng2020inferring} to social science \cite{Pincus2004financial, bury2013market, zeng2014financial}.

Meanwhile, the Prisoner's Dilemma (PD) game has long been studied for the emergence of cooperation among connected (links of the network) selfish individuals (nodes of the network). With the PD game, the criminals could take cooperation (confessing the crime to the judge) or defection (denying the crime) strategy.
A large number of studies have focused on the collective behaviors with different network structures, {\it{e.g.}}, small-world network \cite{Masuda2003SWN}, two-dimensional lattice \cite{Szabo1998squareLattice} and scale-free network \cite{Chen2007scalefree}, etc.. 
Furthermore, the couplings between players are typically symmetric and binary (with / without interactions) between players in those studies.
To quantify the strength of interactions between players, here we studied fully connected asymmetric Sherrington-Kirkpatrick (SK) \cite{kirkpatrick1978infinite} models, with an asymmetric degree parameter could be varying between 0 and 1. Thus the couplings between players are Gaussians and not necessarily symmetric, which are more closer to the realities in natural social systems. We are more interested in the reconstruction of the couplings between players instead of the collective behaviours of the system.


For an evolutionary PD game, the players' strategies follow a binary-state dynamics, which could be naturally mapped into binary state, {\it{i.e.}}, $+1$ for cooperation and $-1$ for defection.
Hence, it is straightforward to pose the coupling inference problem for an evolutionary game into the inverse Ising problem.
There are many related algorithms to recovering the couplings for the inverse Ising problems, as reviewed in \cite{Nguyen2017} and references therein.

Generally, the couplings can be reconstructed by equilibrium \cite{Aurell2012plm, Cocco14058PNAS, Weigt2009PNAS} or non-equilibrium  \cite{pillow2008Nature, Roudi2011PhysRevLett, Zhang2012JSP, zeng2011network,  zeng2013maximum} model given the time-dependent/independent system data. However, in many applications, equilibrium Ising models are typically not the ideal choice for network inference and statistical modeling of the data.
A mismatch between the real couplings and the inferred ones can be seen for large systems \cite{roudi2009pairwise, roudi2011mean, ZengAurell2020review}.
This is mainly because these systems have the out-of-equilibrium nature and  asymmetric couplings between elements.
Thus, researchers have moved to the {\it kinetic Ising models}, developed exact and approximate methods for network reconstruction  \cite{cocco2009neuronal, roudi2011mean, zeng2011network, zeng2013maximum}, which has wider generality.

With kinetic Ising models, spins could update synchronously or asynchronously.
Similarly, with evolutionary games, {\it synchronous} means all players update their strategies simultaneously at every time step \cite{nowak1992evolutionary}.
This can be mathematically described by a finite difference equation, but rarely exists in social or natural systems.
The players, or organisms usually act at different and uncorrelated times with the coming information which may be delayed and imperfect \cite{Huberman1993asynstrategy}.
For instance, a neuron will be in a refractory period after releasing a spike, it cannot respond to an input signal because it is still processing or recovering from the previous input signal \cite{greil2007critical, zeng2020scaling}. The dynamics of such {\it asynchronous update} is usually expressed by differential equations, whose solutions are not always the same with those of their finite-difference counterparts.
Thus players will update their strategies asynchronously here, even though all of them get their corresponding rewards through the game with and from all neighbors. With asynchronous updates of strategies, only one player will be assigned to update the strategy in one game round. For finite number of players $N$ this is not exactly the asynchronous update described above, but we do not see any differences at least for kinetic Ising models \cite{zeng2011network, zeng2013maximum}.

For the network reconstruction with evolutionary game data, some inference methods \cite{wang2016data, Wang2011PhysRevX} have been developed based on compressive sensing technique \cite{foucart2017mathematical}, which is a paradigm for sparse-signal reconstruction.
However, the compressive sensing methodology works with assumptions that the interactions between individuals should be explicit and stationary and the system could be handled linearly in some way.
For example, in the studies of \cite{Wang2011PhysRevX, guo2016roles, wu2016reconstructing}, the observations are the strategies and the total payoff of each player in the evolutionary game.
Furthermore, all agents are assumed to play the same game with a given payoff matrix.
One player's total payoff is hence the sum of the payoff from all neighbors.
In studies~\cite{steinke2007experimental, chang2014exact, Li2017pre}, the dynamical functions are employed, but which can also be decomposed into a linear sum of scalar basis functions.

Be distinct from compressive sensing based inference methods, we present here an exact method which is intuitive and based on maximum-log-likelihood estimation (\textbf{MLE}). Another approximate reconstruction method is presented also which is based on naive mean-field (\textbf{nMF})
approximation.
Both of the derived inference formula for evolutionary PD game data is enlighten by the asynchronously updated Glauber dynamics \cite{glauber1963time}, which could converge to a stationary state described by Gibbs-Boltzman distribution when the couplings are symmetric. However, this is not necessarily true for synchronous case. Hence, we are interested in using a kinetic Ising-like model to reconstruct an evolutionary game network dynamically.

With evolutionary PD game, the update times $\{\tau_i\}$, history of strategies $\{s_i(t)\}$ and rewards for each player $\{x_i(t)\}$ are the full model data generated during  the game process. MLE needs all of these data while nMF does not need the update times, neither the rewards for each game round.  nMF method relies on means, correlations and  the temptation of defection parameter $b$ parameter in the payoff matrix. With massive numerical simulations, we find that the nMF performs very close to MLE method while avoid long-time iteration process. One equilibrium inference method also tested for the sake of completeness, which shows poor reconstruction for the couplings of our non-equilibrium processes.

The paper is organized as follows. In Sec. \ref{sec:structure-dynamics} we present the testing network structure as well as the dynamics for the PD evolutionary game. In Sec. \ref{sec:MLE-nMF}, we derive the MLE and nMF inference formulae from evolutionary game data. Sec. \ref{sec:simulations}, presents how these two methods perform for the network reconstruction, while Sec.
\ref{sec:discussion} summarizes and discusses the results.


\section{Asymmetric SK Model and Glauber-Like Game Dynamics}\label{sec:structure-dynamics}
The quenched aSK model is composed by $N$ vertices, which represents $N$ game player with two kinds of strategies ($s_i=1$ for cooperation while $s_i=-1$ for defection).
This is a fully connected model, \emph{i.e.}, all players are connected to each other. The interactions $J_{ij}$ between each pair of players have the form of
\begin{equation}\label{eq:ask_strcture}
  J_{ij} = J_{ij}^s + k J_{ij}^{as}, ~~~~~~k\ge 0,
\end{equation}
with $k$ the asymmetric degree of these interactions, while $J_{ij}^s = J_{ji}^s$ and $J_{ij}^{as} = - J_{ji}^{as}$ are symmetric and asymmetric matrices, respectively. They are Gaussian random variables with means $0$ and variances
\begin{equation}
  \sigma^2(J_{ij}^s ) =  \sigma^2(J_{ij}^{as} ) = \frac{g^2}{N}\frac{1}{1+k^2}.
\end{equation}
The players do not play game with themselves. 
This indicates $J_{ii} = 0$ and the diagonal elements of $J_{ij}^s$ and $J_{ij}^{as}$ are $0$ as well.

We now describe the dynamics for evolutionary game on the above aSK network model.
Within an evolutionary PD game, the judge knows the players are criminal but without evidences. Then a player has two possible strategies (S): cooperation (confess their  crime to the judge) or defection (deny their offense to the judge). They will get different payoff when they take different strategies.
The rewards are usually expressed by a payoff matrix ${X}$,  for the prisoner dilemma (PD) game:
\begin{equation}\label{eq:payoffM}
X=\left(
\begin{matrix}
    1  & 0  \\
    b  & 0  \\
\end{matrix}
\right),
\end{equation}
with $1\leq b \le 2$. 
If both players choose cooperation, then each of them will get a reward of $1$; if one takes cooperation while the counterpart defects, then the cooperated one will get a reward of $0$ while the defected one gets $b$ (temptation of defection); if both players choose defection, then none of them will get any rewards.

During each time step $t$, all players play games simultaneously and get corresponding rewards according to the payoff matrix $X$ in matrix~(\ref{eq:payoffM}). Thus, for each player, the rewards come form the games played with as well as from the neighbours. Once they get the rewards from the game, one randomly chosen player $i$ 
will be assigned to update his strategy according to the following formula:
\begin{equation}\label{eq:updates}
P(s_i(t)) =\frac{ \exp\left(\sum_j J_{ij}x_j(t-1)\delta_{s_j(t-1), s_i(t)}\right)}{Z_i(t)},
\end{equation}
where $$Z_i(t)=\sum_{s_i(t)=\pm1}\exp\left(\sum_j J_{ij}x_j(t-1)\delta_{s_j(t-1),s_i(t)}\right)$$ is the normalization factor (local partition function), which sums over both cooperation and defection states of player $i$. 

We denote the cooperative and defective strategy  of player $i$ at time $t$ as $\alpha_i(t)$ and $\beta_i(t)$ respectively
\begin{subequations}
    \begin{equation}
        \alpha_i(t) =\sum_j J_{ij}x_j(t-1) \delta_{s_j(t-1),+1} ,
    \end{equation}
    \begin{equation}
       \beta_i(t) =\sum_j J_{ij}x_j(t-1) \delta_{s_j(t-1),-1}.
    \end{equation}
\end{subequations}
Then the partition function can be rewritten as
\begin{equation}\label{eq:Zalphabeta}
  Z_i(t)=\exp\left(\alpha_i(t)\right) + \exp\left(\beta_i(t)\right).
\end{equation}
Define
\begin{eqnarray}\label{eq:external_field}
  H_i(t) &=& \frac{1}{2}\left(\alpha_i(t) - \beta_i(t) \right) \nonumber\\
           &=&\sum_j J_{ij} s_j(t-1)\frac{ x_j(t-1)}{2},
\end{eqnarray}
and
substitute $Z_i(t)$  in~(\ref{eq:updates}) with Eq. (\ref{eq:Zalphabeta}), then we have
\begin{eqnarray}\label{eq:update_rule}
  p(s_i(t)|\mathbf{S})
  &=& \frac{\exp(\alpha_i(t))}{\exp(\alpha_i(t)) + \exp(\beta_i(t))}\nonumber\\
  &=& \frac{\exp(s_i(t)H_i(t))}{2\cosh(H_i(t))}  \nonumber\\
  &=&\frac{1}{2}\left[1+s_i(t)\tanh\left( H_i(t)\right)\right].
\end{eqnarray}

During each time step, all players join the prisoners' dilemma game and obtain corresponding payoff according to the payoff matrix (\ref{eq:payoffM}) from each counterpart. The averaged gain during this time step for player $i$ is derived in Eq. (\ref{eq:xt}).
However, only one of the players is going to update his strategy, which is not necessarily change his current choice. This is why we refer this dynamics as {\it asynchronous} PD game.

\section{INFERENCES FOR GAME NETWORKS}\label{sec:MLE-nMF}
\subsection{Maximum Log-likelihood Estimation (MSE)}
Suppose we have the following data: history of players' strategies $\textbf{s}\equiv\{s_i(t)\}$,  updating times $\mathbf{\tau} \equiv \{\tau_i\}$ and payoff of each player $\textbf{x}\equiv \{x_i(t)\}$, of length $L = T\times N$ steps.
We reconstruct the game network based on these data.
This can be done by maximizing the likelihood $P(\textbf{s},\mathbf{\tau},\textbf{x})=P(\textbf{s},\textbf{x}|\mathbf{\tau})p(\mathbf{\tau})$ over these parameters. As mentioned in \cite{zeng2013maximum}, for each player $i$, the chance of updating their strategy is an Poisson process.
This indicates the probability of the update history $p(\mathbf{\tau})$ is independent of the model parameters, thus we can take the objective function as $\log P(\textbf{s},\textbf{x}|\mathbf{\tau})$ safely from Eq. (\ref{eq:update_rule}),
\begin{equation}\label{eq:loglikelihood}
 \mathcal{L}=\sum_i\sum_{\tau_i}\left[s_i(\tau_i+\delta t)H_i(\tau_i)-\log 2\cosh H_i(\tau_i)\right].
\end{equation}
The sum only over the update times instead of all time steps as in synchronous case \cite{roudi2011mean}.
Then it leads to a learning rule for couplings $J_{ij}$
\begin{eqnarray}\label{eq:MLE-J}
 \delta J_{ij} &\propto& \frac{\partial{\mathcal{L}}}{\partial{J_{ij}}} \nonumber \\
  &\propto&  \sum_{\tau_i}\left[ s_i(\tau_i+\delta t)-\tanh H_i(\tau_i)\right]s_j(\tau_i)\frac{x_j(\tau_i)}{2}.~~~~
\end{eqnarray}
We refer this algorithm as the ``\MLE''.
It is an exact method as no approximation is introduced to the derivation of the inference formula. Besides, it is notable that if the rewards for each player equal to $2$ and the same with each other, then the PD game process degenerates to the original Glauber dynamics \cite{glauber1963time}.

\subsection{naive Mean Field (nMF) approximation}

We start from the master equation
\begin{eqnarray}\label{eq:master-equ}
 \frac{dp(\mathbf{S}; t)}{dt} &=&\sum_i\omega_i(-s_i(t)) p(s_1,...,-s_i,...s_N;t) \nonumber \\
 &-&\sum_i\omega_i(s_i(t))p(\mathbf{S};t),
\end{eqnarray}
and the flipping rate is:
\begin{equation}\label{eq:flipping-rate}
 \omega_i(\mathbf{S};t) = \frac{1}{2}\left[1-s_i(t)\tanh\left(H_i(t)\right)\right].
\end{equation}
which could be obtained through detailed balanced condition \cite{van1992stochastic}:
\begin{subequations}
\begin{equation}  \frac{p(-s_i)}{p(s_i)}=\frac{\omega(s_i)}{\omega(-s_i)}
   \end{equation}
   \begin{equation}
  p(-s_i) + p(s_i) = 1
  \end{equation}
\end{subequations}

With strategy history for each player $\{s_i(t)\}$, the time-dependent means and correlations can be defined as
\begin{subequations}
  \begin{equation}\label{eq:mean-correlation}
 m_i(t) = \langle s_i(t) \rangle,~~~~~~~~~~~~~~~~~
 ~~~~~~~~
 \end{equation}
\begin{equation}
 C_{ij}(t-t_0) = \langle s_i(t)s_j(t_0) \rangle -m_i (t)m_j(t).
\end{equation}
\end{subequations}

With Eqs. (\ref{eq:master-equ}) and (\ref{eq:flipping-rate}), the temporal behavior of means and correlations are
\begin{subequations}
 \begin{equation}\label{eq:time-derivative-MC}
  \frac{dm_i(t)}{dt} = -m_i(t) + \langle \tanh[H_i(t)] \rangle,~~
  \end{equation}
  \begin{equation}\label{eq:time-derivative-MC-2}
  \frac{d\langle s_i(t)s_j(t_0)\rangle}{dt} = -\langle s_i(t)s_j(t_0) \rangle + \langle\tanh[H_i(t)]s_j(t_0) \rangle.
 \end{equation}
\end{subequations}

Thus the external field (influence) $H_i(t)$  for player $i$  at time $t$  in (\ref{eq:external_field}) could be rewritten as
 \begin{eqnarray}\label{eq:Hit}
  H_i(t) &=& \sum_j J_{ij}s_j(t)x_j(t) \nonumber \\
  &=& \sum_j J_{ij}\left(B_0 + B_1 s_j(t-1)\right),
 \end{eqnarray}
 where $B_0=(1-b)(1+\bar{m}(t))/4$ and $B_1=(1+b)(1+\bar{m}(t))/4$ are two non-fluctuating term depending on parameter $b$ and $\bar{m}(t)=\frac{1}{N}\sum_i s_i(t)$ is the population average.
 The detailed derivation of Eq. (\ref{eq:Hit}) is shown in Appendix \ref{app:external-field}.

The nMF approximation hence is formally the same with that for original asynchronously updated kinetic Ising model in \cite{zeng2011network},
\begin{equation}
  m_i = \tanh b_i,
\end{equation}
but with
\begin{equation}
b_i = \sum_{j \in \partial i} J_{ij}(B_0 + B_1m_j),
\end{equation}
which is obtained by taking the stationary solution of Eq. (\ref{eq:time-derivative-MC})  with naive mean field approximation. The details are illustrated in Appendix \ref{app:nMF}.

With the time-derivative of correlations in Eq. (\ref{eq:time-derivative-MC-2}), the $\tanh$ function can be expanded with respect to $b_i$, yields
\begin{eqnarray}
&& \frac{d}{dt}\left<s_i(t)s_j(t_0)\right>
 +\left<s_i(t)s_j(t_0)\right>\\ &=&  m_i m_j+B_1(1-m_i^2)
\sum_kJ_{ik}\left<\delta s_k(t)\delta s_j(t_0)\right>.
\end{eqnarray}
Denoting $\tau = t-t_0$, we get
\begin{equation}
 \frac{d}{d\tau}C_{ij}(\tau) + C_{ij}(\tau) = B_1(1-m_i^2) \sum_k J_{ik} C_{kj}(\tau),
\end{equation}
With the limit $\tau\rightarrow 0$, we get the \nMF~ inference formula for the PD game network
\begin{eqnarray}\label{eq:nMF-J}
 J &=& \frac{A^{-1}DC^{-1}}{B_1},
\end{eqnarray}
with $D=\dot{C}(0)+C(0)$, $A_{ij} = \delta_{ij}(1-m_i^2)$ and $B_1=(1+b)(\bar{m}+1)/4$.
With full data history, it is safe to replace $\bar{m}(t)$ with the time average $\bar{m}=\left<\bar{m}(t)\right>_{t} $.

\section{Simulation Results}\label{sec:simulations}
\begin{figure}[!h]
\centering
 \includegraphics[width=0.49\textwidth]{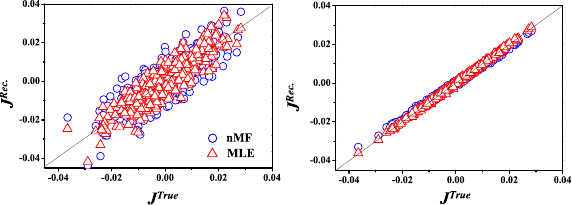}
 \put(-225,80){(a)}
 \put(-100,80){(b)}
 \caption{Scatter-plots for reconstructed couplings versus the tested ones with different data length $L$. Blue open dots are recovered couplings by nMF while open red upper triangles are for MLE. Left panel: data length $L=N\times 10^5$; right panel: data length $L=N\times 10^7$. The other parameters are the same: number of players $N=20$, asymmetric degree $k=1$, coupling strength $g=0.3$. }\label{fig:scaterJs}
 \end{figure}

We infer the evolutionary game network by MLE (\ref{eq:MLE-J}) and nMF (\ref{eq:nMF-J}) formula respectively. In Fig. \ref{fig:scaterJs}, the scatter plots for reconstructed structure against the true ones are presented for both methods. Fig.~\ref{fig:scaterJs}(a) is for the data length $L=N\times10^5$ and Fig.~\ref{fig:scaterJs}(b) for $L=N\times10^7$.
It shows that both nMF and MLE perform better with larger data length $L$. Furthermore, MLE works slightly better than nMF in both cases. The nMF method over estimates the couplings especially for $L=N\times10^7$, as shown in Fig. \ref{fig:scaterJs}(b).

We introduce the mean square error (MSE) to measure the distance between the reconstructed and the true tested network structure.
\begin{equation} \label{eq:MSE}
 \text{MSE} = \frac{\sum_{ij}(J_{ij}^{Rec.}-J_{ij}^{True})^2}{N(N-1)},
\end{equation}
where $J_{ij}^{True}$ are the tested network couplings and $J_{ij}^{Rec.}$ for the inferred ones.

With the reconstruction error MSE defined by (\ref{eq:MSE}), we compared the performance of MLE and nMF method for fully connected SK model (\ref{eq:ask_strcture}).
We study both methods for different data length $L$, the asymmetric degree $k$, the temptation of defection parameter $b$  in the payoff matrix (\ref{eq:payoffM}) as well as the coupling strength $g$.
The inverse of $g$ corresponds to the temperature in spin glass system, which can be explained as the noise (say, fake information or rumors etcetera) in the evolutionary PD game.

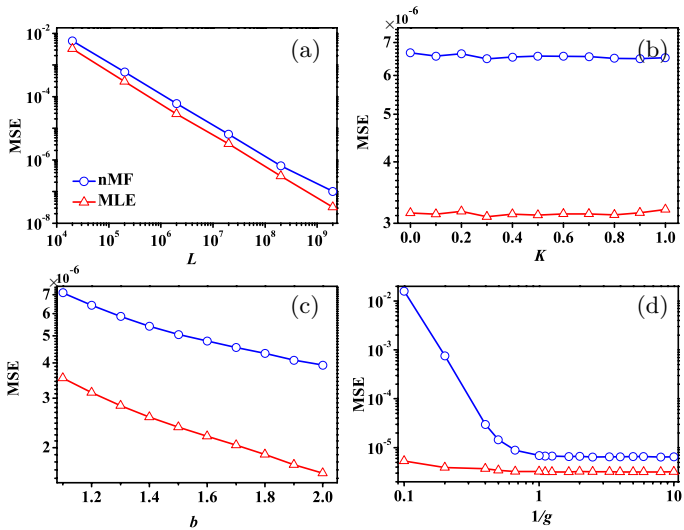
\begin{figure}[!h]
\centering
 \includegraphics[width=0.5\textwidth]{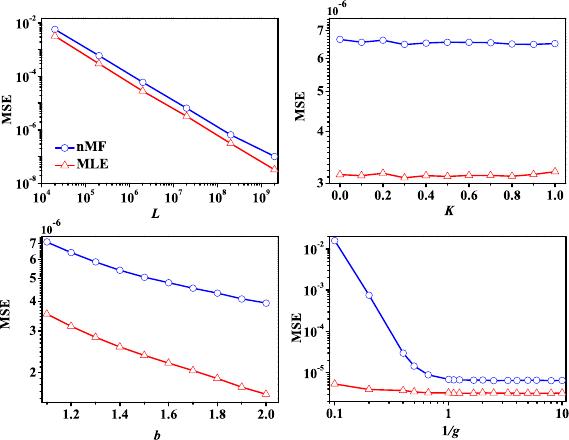}
 \put(-240,91.5){\tiny $\times$}
 \put(-113,189){\tiny $\times$}
 \put(-148,178){(a)}
 \put(-17,178){(b)}
 \put(-148,81){(c)}
 \put(-17,81){(d)}
 \caption{(color online). Mean square error (MSE) for (a) data length $L$ , (b) asymmetric degree $k$, (c) $b$ parameter in payoff matrix and (d) temperature $1/g$ respectively. Blue open dots present nMF while red solid dots for MLE. The parameters are $N=20$, $L=N\times10^6$, $k=1$ and $g=0.3$ except varied in the corresponding panel.}
 \label{fig:mses}
 \end{figure}

Figure \ref{fig:mses} shows the performance of both methods.
Similar to the results for asynchronous Ising case \cite{zeng2013maximum}, the MSE for MLE~is $\propto 1/L$ here.
However, with the data length $L$ we tested here, nMF does not approach to a saturated value as happened for asynchronous Ising case.
Instead, the reconstruction error MSE for nMF is about twice of that for MLE, see Fig. \ref{fig:mses}(a).
This could be an indication for game network, nMF inference formula can be obtained also by taken the average of update times in MLE formula (\ref{eq:MLE-J}), which has been proved so in \cite{zeng2013maximum} for asynchronous Ising models.
Figure \ref{fig:mses}(b) shows the MSE for both methods is not sensitive to the asymmetric degree $k$, no matter the network is symmetric ($k=0$) or fully asymmetric ($k=1$). Here, the data length $L=N \times 10^6$ with $N=20$ and $g=0.3$.  The difference between two methods is effected mainly by the data length $L$.
The MSE for MLE is about half of that for nMF approach, which agrees well with that shown in Fig. \ref{fig:mses}(a).
Figure \ref{fig:mses}(c) shows the performance of MLE and nMF algorithm for different reward parameter $b$ in the payoff matrix (\ref{eq:payoffM}), which is a specific parameter for game networks. Both algorithms recover the network better with larger reward $b$. However, the proportion of MSE between nMF and MLE increases slightly for larger $b$. Finally, the effects of inverse $g$ over network inference is illustrated in Fig. \ref{fig:mses}(d). For strong couplings, nMF works much worse than MLE. However, for $g\le 1$, their performances are not effected by $g$ anymore. MLE still has its advantage of the factor 2 (conferred by the information of update times) compared with nMF.

\begin{figure}[!h]
\centering
 \includegraphics[width=0.49\textwidth]{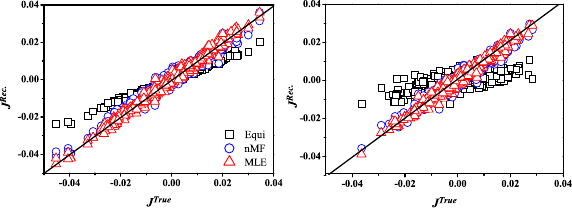}
 \put(-225,80){(a)}
 \put(-100,80){(b)}
 \caption{Scatter plot for recovered links with the tested ones with different asymmetric degree $k$. Black open squares are for inferred couplings with equilibrium inference $J^{equi} = -C(0)^{-1}$. Blue open dots are for asynchronous nMF while  red open triangles are for MLE. Left panel: asymmetric degree $k=0$, symmetric SK model; right panel: $k=1$, fully asymmetric SK model. The other parameters are the same: data length $L=N\times 10^6$, number of players $N=20$, coupling strength $g=0.3$. }\label{fig:scaterJ-ks}
 \end{figure}

To illustrate how the equilibrium inference method works for the non-equilibrium process, we use the naive mean-filed equilibrium inference \cite{Kappen2000} algorithm which depends on the equal time correlations only:
 \begin{equation}\label{eq:equinMF}
  J^{Equi} = -C^{-1}.
\end{equation}
We refer this as equilibrium inference to avoid the confusion with the nMF method for asynchronous case we derived here.
Then, we present  the inferred networks through scatter-plots in Fig. \ref{fig:scaterJ-ks} with both equilibrium and non-equilibrium method for evolutionary data generated by PD games.

Fig. \ref{fig:scaterJ-ks}(a) and (b) presents the reconstructed couplings for symmetric ($k=0$) and fully asymmetric ($k=1$) SK model respectively. Our asynchronous nMF (blue open dots) and MLE (red upper triangle) methods recover the couplings nicely while the equilibrium nMF (\ref{eq:equinMF}) (open black squares) works much worse, even for $k=0$, the symmetric case. It is not surprising as equilibrium nMF (\ref{eq:equinMF}) we tested here does not take into account the game process at all. 
For the original asynchronous Ising model without game process, the equilibrium nMF should work as good as the non-equilibrium methods \cite{ZengAurell2020review} since they are for the same dynamics.

\section{Discussion}\label{sec:discussion}
We studied the network reconstruction problem with asynchronous updated evolutionary game data. The
gamblers play the PD game with their counterparts simultaneously but update their strategies asynchronously.     Two methods (exact and iterative MLE and approximate nMF) are introduced to infer the fully connected and Gaussian distributed couplings between players.

Comparing with the original asynchronous Ising model  \cite{zeng2011network, zeng2013maximum}, we combine the PD games with Glauber dynamics, where the rewards during the game for each player is considered.
The MLE and nMF reconstruction formulae for asynchronous Ising models in \cite{zeng2011network, zeng2013maximum} hence have to be generalized for the asynchronous evolutionary game case here.
MLE is derived from the log-likelihood of the system
while nMF from the derivative of the equations of motion of means and correlations.
However, they are not independent with each other.
MLE utilizes the full model histories, $\{s_i(t)\}$, $\{x_i(t)\}$ and $\{\tau_i\}$. The update histories in Eq. (\ref{eq:MLE-J}) could be averaged and yields the nMF inference formula (\ref{eq:nMF-J}) by taking the linear term of the averaged $\tanh$ function. 

With the derived MLE and nMF methods for asynchronously updates PD game, our numerical results show that the inference performance of nMF is comparable with that of MLE.
Both methods are not sensitive to the asymmetric degree $k$, but proportional to the inverse of data length. For fixed data length, the reconstruction error (MSE) of MLE is half (benefit from update times) of that by nMF method.
Furthermore, nMF prefers small/weak coupling strength ($g$) while MLE holds for much wider range of $g$. Both methods recover better for large payoff parameter $b$. This is reasonable as when the rewards of each players equal to $2$, then the external field Eq. (\ref{eq:external_field})  in the dynamics degenerates to the original asynchronous Ising model.
With $b$ increases, the rewards of each players will also close to $2$ which show better performance of network reconstruction.

However, comparing with the data that MLE method used for inference, nMF needs only the $b$ parameter and the moments of strategy history $\{s_i(t)\}$. Furthermore, nMF infers much fast than MLE as no iterative procedure included for network inference. 
The scatter-plots in Fig. \ref{fig:scaterJs} and  Fig. \ref{fig:scaterJ-ks} show that the $J_{ij}^{nMF}$s are obviously comparable with $J_{ij}^{MLE}$s, even though the MSE for nMF is roughly twice of that for MLE.
Thus, we expect that for the network inference with data from asynchronous PD game, nMF works well enough. 
The higher orders of tanh function (say, TAP) are not necessary.

Besides, we have tested a widely used equilibrium inference method (\ref{eq:equinMF}) blindly for our dynamic data generated from evolutionary PD games, which recover much worse than that by non-equilibrium inference methods, even for symmetric interactions between players, as illustrated in Fig. \ref{fig:scaterJ-ks}(a).
This bring us a message that if we have time series data generated from some certain dynamics, we should try non-equilibrium methods first. Furthermore, we start from the detailed balance condition to derive the asynchronous nMF formula for the network inference, but as shown by the numerical results, the inference formula holds for asymmetric $k=1$ case also where the system is not in the equilibrium realm any more. 

This work focuses on a simple and idealized prisoner's game with fully connected and Gaussian distributed interactions, which could be extended to certain type of structures or different distributions (say, random power-law, where strong couplings between few players while weak ones between most of them) for the couplings. Besides, we use a  constant value of defection temptation $b$ which could be different for each players. Meanwhile, non-zero abut small $r$ in the payoff matrix (\ref{eq:payoffM}) could also be tested. These will be studied by further work.

\appendix
\section{Reformulating the external field $H_i(t)$ } \label{app:external-field}
According to the payoff matrix (\ref{eq:payoffM}) for prisoner's dilemma game, the average reward that player $i$ will gain at time $t$ is (for simplicity, $r$ is assumed to be equal to 0):
\begin{eqnarray}\label{eq:xt}
  x_i(t) &=& 2\times \left\{ \frac{1+s_i(t)}{2}\frac{1}{N}\sum_{j\in\partial i}\frac{1+s_j(t)}{2}\right.   \nonumber  \\
&+& \left. \frac{1-s_i(t)}{2}\frac{b}{N}\sum_{j\in\partial i}\frac{1+s_j(t)}{2}\right\}.
\end{eqnarray}
The factor $2$ outside the bracket comes from the fact that players obtained rewards from the game with neighbors as well as the games originated from neighbors within each time step of synchronous PD games.
Then $s_i(t)\frac{x_i(t)}{2}$ in Eq. (\ref{eq:external_field}) is:
\begin{eqnarray}\label{eq:xs-0}
s_i(t)\frac{x_i(t)}{2} &=& \frac{s_i(t)+1}{2}f(c)
 +\frac{s_i(t)-1}{2}bf(c),
 \end{eqnarray}
with $f(c)$ the cooperation density in the neighborhood of player $i$:
\begin{equation}
 f(c)=\frac{1}{N} \sum_{j\in\partial{i}} \frac{1+s_j(t)}{2}=\frac{\bar{m}(t)+1}{2}.
\end{equation}
Hence we rewrite Eq. (\ref{eq:xs-0}) as:
\begin{equation}\label{eq:xs}
s_i(t)\frac{x_i(t)}{2}= B_0 + B_1 s_i(t),
 \end{equation}
with
\begin{subequations}
  \begin{equation}\label{eq:B0}
 B_0 = \frac{1-b}{2}f(c)= \frac{(1-b)(\bar{m}(t)+1)}{4},
\end{equation}
\begin{equation}\label{eq:B1}
 B_1 = \frac{1+b}{2}f(c) = \frac{(1+b)(\bar{m}(t)+1)}{4},
\end{equation}
\end{subequations}
and $\bar{m}(t) = \frac{1}{N}\sum_i s_i(t)$.

Thus, we reformulated eq. (\ref{eq:external_field}) as:
\begin{equation}
  H_i(t) = \sum_j J_{ij}\left(B_0 + B_1s_j(t-1)\right).
\end{equation}

\section{nMF approximation}\label{app:nMF}
Substituting $s_j(t-1)$ with non-fluctuating and fluctuating terms as:
\begin{equation}
 H_i(t) = \sum_j J_{ij}(B_0 + B_1 (m_j + \delta s_j(t-1))),
\end{equation}
and define
\begin{equation}
  b_i=\sum_j J_{ij}(B_0+B_1 m_j),
\end{equation}
 then
\begin{eqnarray}
\left<\tanh[H_i(t)]\right> &=& \left<\tanh\left[b_i + B_1\sum_j J_{ij}\delta s_j(t-1) \right]\right>.\nonumber\\
\end{eqnarray}
expanding the $\tanh$ function with respect to $b_i$ to the first order, we get:
\begin{eqnarray}
&&\left<\tanh[H_i(t)]\right> \nonumber\\
&=&  \tanh b_i + B_1(1-\tanh^2 b_i)\left< \sum_j J_{ij}\delta s_j(t-1)\right>,~~~~~~  
\end{eqnarray}
where the second term on the right hand-side goes to zero. Hence we have naive mean-field approximation for $m_i$ as:
\begin{equation}\label{eq:mi}
 m_i = \tanh b_i.
\end{equation}
which is the stationary solution of the time derivative of magnetization in equation (\ref{eq:time-derivative-MC}).

\begin{acknowledgments}
We sincerely thank John Hertz, Erik Aurell and Pan Zhang for heuristic and fruitful discussions and suggestions.
H.-L. Zeng was supported by NSFC Project No. 11705097.
HLZ was also supported by Natural Science Foundation of Jiangsu Province (BK20170895),
Natural Science Fund for Colleges and Universities in Jiangsu Province (17KJB140015),
Jiangsu Government Scholarship for Overseas Studies of 2018, and
Scientific Research Foundation of Nanjing University of Posts and Telecommunications (NY217013).
S.-M. Qin was supported by Project NSFC No. 11705279.
\end{acknowledgments}

\bibliography{./inverseIsing_PDgame}
\end{document}